\documentstyle[epsf]{elsart}
\begin{document}

\def\al{\alpha}
\def\be{\beta}
\def\de{\delta}
\def\eps{\epsilon}
\def\e{{\rm e}}
\def\sech{{\rm sech}}
\def\ka{\kappa}
\def\ga{\gamma}
\def\om{\omega}
\def\Om{\Omega}
\def\th{\theta}
\def\ze{\zeta}
\def\et{\eta}
\def\vph{\varphi}
\def\si{\sigma}
\def\fr{\frac}
\def\ov{\over}
\def\pd{\partial}
\def\cd{\cdot}
\def\ra{\rightarrow}
\def\lra{\leftrightarrow}
\def\fr{\frac}
\def\bk{\mbox{\boldmath $k$}}
\def\bx{\mbox{\boldmath $x$}}
\def\beq{\begin{equation}}
\def\eeq{\end{equation}}
\def\beqr{\begin{eqnarray}}
\def\eeqr{\end{eqnarray}}
\def\nn{\nonumber}

\begin{frontmatter}

\title{Complexation of Wavenumbers in Solitons}
\author{M. Umeki}
\footnote{Permanent address: 
Department of Physics, University of Tokyo, 
7-3-1, Hongo, Bunkyo-ku, Tokyo, 113 Japan. \ 
E-mail: umeki@newton.Colorado.EDU}
\address{
Department of Applied Mathematics,
University of Colorado at Boulder, 
Boulder, CO 80309-0526 U.S.A.}

\begin{abstract}
It is shown that, by letting wavenumbers and frequencies complex 
in Hirota's bilinear method, new classes of exact solutions 
of soliton equations can be obtained systematically. 
They include not only singular or 
$N$-homoclinic solutions but also $N$-wavepacket solutions. 
\end{abstract}

\end{frontmatter}

{\small 
}

Recent awareness of homoclinic solutions of soliton equations
\cite{Herbst,Umeki} 
has made it possible to apply them for studies of chaotic 
phenomena in nonlinear, infinite dynamical and complex 
systems, e.g., \cite{Bishop}. 
In contrast to the low-dimensional ordinary differential 
equations, partial differential equations may have 
multiple or infinite homoclinic structures. 
Therefore, it is important to find and classify 
homoclinic solutions systematically. 

The key idea of this paper is to make wavenumbers complex 
in soliton solutions. Usually this procedure fails because 
the dependent variable becomes complex for real-valued 
equations. 
However, if we keep the complex conjugate pair for wavenumbers, 
the difficulty can be avoided.

Let us start with the simplest system, the KdV equation,
\beq
 u_t + 6 u u_x + u_{xxx}=0, 
\label{a1}
\eeq
where $u_x = \pd_x u = \pd u/\pd x$. 
Letting $ u(x,t) = 2 ( \ln f(x,t) )_{xx},$ the bilinear form becomes
\cite{Hirota1}
\beq
M f \cd f=0, 
\label{a2}
\eeq
where 
\beq
M=M(D_t, D_x)= D_x(D_t+D_x^3),
\label{a3}
\eeq
and $D_x$ is the Hirota's bilinear operator,
\beq
D_x f \cd g = (\pd_x - \pd_{x'}) f(x)g(x')|_{x'=x}. 
\label{a4}
\eeq

The $N$-soliton solution is given by
\beq
f=\sum_{\mu = 0,1} \exp \left( \sum_{j=1}^N \mu_j \eta_j + 
\sum_{j>l}^N \mu_j \mu_l A_{jl} \right) , 
\label{a5}
\eeq
with
\beq
\eta_j = k_j x + \Om_j t +\eta_{0j}, 
\label{a6}
\eeq
\beq
\Om_j = -k_j^3, 
\label{a7}
\eeq
\beq
\exp A_{jl} \equiv a_{jl} = -\frac{M(\Om_j-\Om_l,k_j-k_l)}
{M(\Om_j+\Om_l,k_j+k_l)},
\label{a8}
\eeq
where $\sum_{\mu = 0,1}$ indicates the summation over all 
possible combinations of $\mu_i = 0,1$ for $i=1, \cdots , N$.

This solution gives solitons for real $k_j$. 
If we make $k_j$ complex conjugate as $k_{j-1}=k_j^*$
for $j=2, 4, \cdots, 2M$ and real for $j=2M+1, \cdots, N$, 
it no longer gives classical solitons. 
We note that the proof of the $N$-soliton solutions by 
mathematical induction \cite{Hirota1} 
is still valid for complex wavenumbers.  

For $M=1$ and $L\equiv N-2M=0$, the function $f$ and 
the solution can be described by 
\beq
f=1+ 2 \exp \xi_1 \cos \ze_1 + a_{12} \exp (2\xi_1), 
\label{a9}
\eeq
\beq
u=4K_1^2 \fr{\e^{-\xi_1} \cos (\ze_1+2 \vph_1)+ 
a_{12}\e^{\xi_1} \cos (\ze_1-2 \vph_1) 
+ 2[a_{12}\cos^2 \vph_1-\sin^2 \vph_1]}
{(\e^{-\xi_1}+a_{12}\e^{\xi_1}+2 \cos \ze_1)^2},
\label{a10}
\eeq
where $\xi_i$ and $\ze_i$ are the real and imaginary parts 
of $\et_i$, and 
\beq
a_{12}= \left(\fr{k_1-k_2}{k_1+k_2}\right)^2 = -\tan^2 \vph_1,
\label{a11}
\eeq
for 
$k_1=K_1 \exp{i \vph_1}.$
Since $a_{12}$ is negative, it can be shown that 
this solution possesses one singular point ($f=0$) 
in $x$ for given $t$. 
Figure 1 shows the 3D plot of this solution. 

\begin{figure}[t]
\begin{center}
\mbox{
\epsfbox{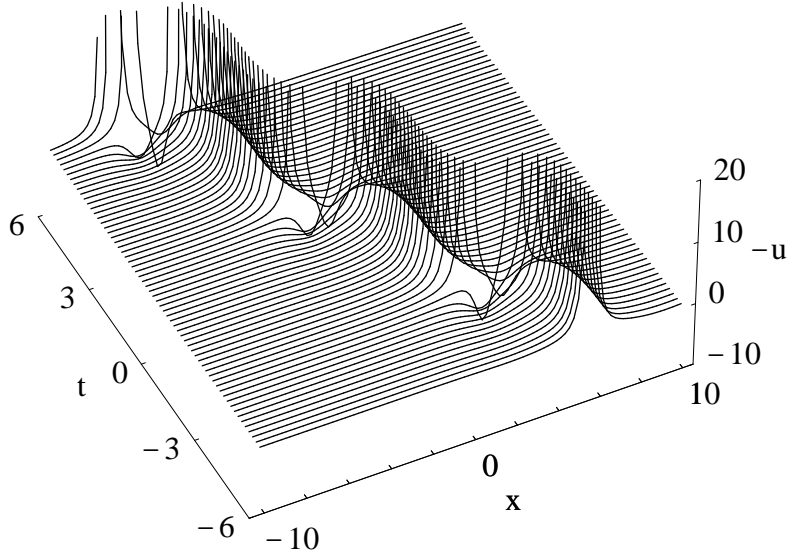}}
\end{center}
\caption{ The 3D plot of a one-singular-point solution 
of the KdV equation with $k_1=\exp (i \pi /4)$. 
The constants $\eta_{i0}$ are zero throughout 
the illustrative examples in this paper.}
\label{fig1}
\end{figure}

It looks as the singular point travels at the velocity 
$-\Om_{1r}/k_{1r}$ (subscript $r$ denotes the real part) 
with oscillating back and forth 
in the stripe surrounded by lines 
$\xi_1=[ \ln (\sqrt{1-a_{12}})\pm 1]/(-a_{12})$, 
which are derived by $f=0$ and $\cos \ze_1=\pm 1 $
in (\ref{a9}).
The singularity behaves as $u\sim u_s = -2(x-x_0)^{-2}$, 
which is derived by a simple zero of $f$ in $x$. 
It should be noted that $u_s$ is an exact steady 
solution of the KdV equation.
It is also observed that a small pulse with 
the sign opposite to the singularity is travelling with 
the velocity $-\Om_{1i}/k_{1i}$.

\begin{figure}[t]
\begin{center}
\mbox{
\epsfbox{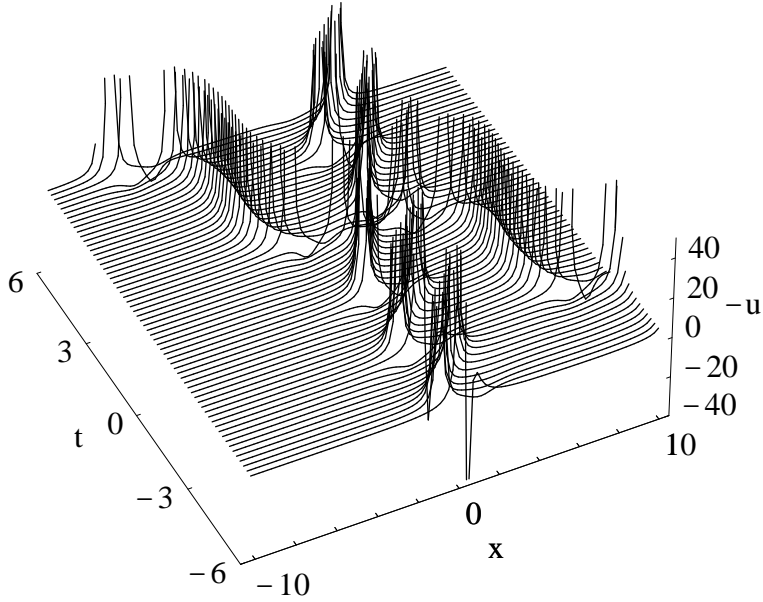}}
\end{center}
\caption{ The 3D plot of the interaction between 
two singular points of the KdV equation 
with $k_{1} = \exp (i \pi /4)$ and $k_{2} = 2^{1/2} 
\exp (i \pi /6)$.}
\label{fig2}
\end{figure}

The interaction between two singular points for $(M,L)=(2,0)$ 
is illustrated in Figure 2. 
The asymptotic behavior of the singular point for $k_1$ 
can be given as follows. In the moving frame $X_1=x+(\Om_{1r}/k_{1r})t$, 
$\et_3$ can be written as 
\beq
\et_3 = k_{3r} X_1+ \si_{3r} t + i (k_{3i} X_1 + \si_{3i} t)+\et_{30}, 
\label{a12} \eeq
where $\si_{3r}=\Om_{3r}-k_{3r}\Om_{1r}/k_{1r}$ and 
$\si_{3i}=\Om_{3i}-k_{3i}\Om_{1r}/k_{1r}$. If $\si_{3r}<0$, 
\beq
f \sim 1+ \e^{\et_1} + \e^{\et_2} + a_{12} \e^{\et_1 + \et_2} =0,
\label{a13} \eeq
as $ t \ra \infty$. Similarly, in the moving frame 
$X_3=x+(\Om_{3r}/k_{3r})t$, 
\beq
\et_1 = k_{1r} X_3+ \si_{1r} t + i (k_{1i} X_3 + \si_{1i} t)+\et_{10}, 
\label{a14} \eeq
where $\si_{1r}=-k_{1r} \si_{3r}/k_{3r}$ and 
$\si_{1i}=\Om_{1i}-k_{1i} \Om_{3r}/k_{3r} $. 
Assuming $k_{1r}/k_{3r}>0$, we have $\si_{1r}>0$ and 
\beq
f \sim a_{12} \e^{\et_1+\et_2} [ 1+ \e^{\et_3+\de_3} 
+ \e^{\et_4+\de_4} + a_{34} \e^{\et_3+\et_4+\de_3+\de_4} ],
\label{a15} \eeq
as $ t \ra \infty$, where the {\it complex} 
phase shift is given by 
$\de_3=\de_4^*=\ln(a_{13}a_{23})$. 

In the limit $t \ra -\infty$ and with the same assumption as the above,
the moving frames $X_1$ and $X_3$ lead to 
the asymptotic forms (\ref{a15}) and (\ref{a13}) with replacement 
$(1,2) \lra (3,4)$ and 
%\beq
%f \sim a_{34} \e^{\et_3+\et_4} [ 1+ \e^{\et_1+\de_1} 
%+ \e^{\et_2+\de_2} + a_{12} \e^{\et_1+\et_2+\de_1+\de_2} ],
%\label{a16} \eeq
the phase shift $ \de_{1}=\de_{2}^*=\ln (a_{13} a_{14})$. 
From the definition of $a_{ij}$, we have $\de_1=\de_3$ and 
the conservation law of the total phase shift holds, i.e., 
$\de_1+\de_2=\de_3+\de_4.$
The conservation law does not depend on the signs of 
$k_{1r}/k_{3r}$ and $\si_{1r}$. 

Figure 3 shows the interaction between a 
singular point and a soliton for $(M, L)=(1,1)$. 
\begin{figure}[t]
\begin{center}
\mbox{
\epsfbox{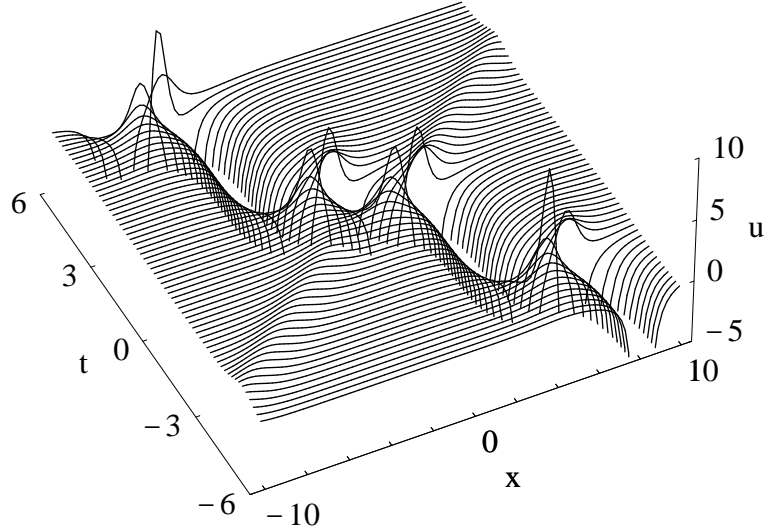}}
\end{center}
\caption{ The 3D plot of the interaction between 
a singular point and a soliton of the 
KdV equation 
with $k_{1} = \exp (i \pi /4)$ and $k_{2} = 2^{1/2}$.}
\label{fig3}
\end{figure}
Under the same assumptions of the signs of 
$k_{1r}/k_{3r}$ and $\si_{1r}$, the asymptotic behavior of 
the soliton is
\beq
u \sim \fr{k_{3r}^2}{2} \sech^2 \fr{\xi_3}{2},
\label{a17} \eeq
as $t \ra -\infty $ and 
\beq
u \sim \fr{k_{3r}^2}{2} \sech^2 \left( \fr{\xi_3}{2}+\de_{3}
\right),
\label{a18} \eeq
as $t \ra \infty $, where the phase shift is given by 
$\de_{3} = \ln (a_{13} a_{23})$. 
Similarly, the behavior of the singular point is determined by the 
zero of (\ref{a13}) and (\ref{a15}) with $(1,2) \lra (3,4)$, 
the factor in front of the square bracket in (\ref{a15}) 
replaced by $\e^{\et_3}$, 
and the phase shift $\de_{1} = \de_2^* = \ln a_{13}$ as 
$t \ra \infty$ and $t \ra -\infty$, respectively. 
Again, the total phase shift is conserved.

As we see in the above examples, the singular points 
behave as if they keep their identities after they collide with 
each other like solitons. Therefore, they may be called {\it singulons}. 

At first sight, it seems that this singularity makes 
the solutions physically meaningless. 
However, the singularity shown above does not 
always appear. 
As such an example, the Boussinesq equation of the form
\beq
u_{tt}-u_{xx}-3(u^2)_{xx}-u_{4x}=0, 
\label{a19}
\eeq
is considered. Substituting $u=2 (\ln f)_{xx}$, 
the bilinear equation
\cite{Hirota2} 
is given by (\ref{a2}) with the operator 
\beq
M=D_t^2-D_x^2-D_x^4. 
\label{a20}
\eeq
The solution is given by (\ref{a5}) with the dispersion relation 
\beq
\Om_i = \pm k_i \sqrt{1+k_i^2}. 
\label{a21}
\eeq

We note that this solution with complex wavenumbers 
corresponds to a certain limit of the periodic solutions by 
two-dimensional $\theta$-function (3.1) of \cite{Nakamura} 
with $\eta_1=\eta_2^*$, although 
it is implicit that the wavenumber and the frequency 
may be complex in \cite{Nakamura}.

If we choose pure imaginary wavenumbers $k_{j-1}=-k_j=iK_{j-1}$ 
and $K_j>1$ for $j=2, \cdots , 2M$ and $L=0$, 
we obtain so-called homoclinic solutions, 
analogous to homoclinic orbits in ordinary differential equations. 
Strictly speaking, they are temporally homoclinic and 
spatially periodic (or quasi-periodic) solutions. 

As an example, let us consider the simplest complexation, 
i.e. $(M,L)=(1,0)$. 
The asymptotic behaviors of the solution given by (\ref{a10}) 
are
\beq
u \sim 4K_1^2 \exp \xi_1 \cos (\ze_1+2 \vph_1),
\label{a22}\eeq
as $t \ra -\infty$ and 
\beq
u \sim 4K_1^2/a_{12} \exp (-\xi_1) \cos (\ze_1-2 \vph_1),
\label{a23}\eeq
as $t \ra \infty$, assuming that $\Om_{1r}$ is positive. 

If we put $k_1=-k_2=i K_{1}$ and $K_{1}>1$, the frequency 
$\Om_j$ becomes real and can be chosen as 
$\Om_1=\Om_2=K_{1}\sqrt{K_{1}^2-1}$. 
In this case, the interaction coefficient is 
$a_{12}=(4K_1^2-1)/(K_1^2-1)$ and positive. 
Since $a_{12}>4$ and $f$ can be written as 
\beq 
f=2\sqrt{a_{12}} \cosh (\xi_1+\de_1) + 2 \cos \ze_1, 
\label{a24}
\eeq
with $\de_1= (\ln a_{12}) /2$, this solution remains 
finite for all $x$ and $t$.

As $t \rightarrow - \infty$, the solution 
tends to a homoclinic point $u=0$ and behaves 
as $u\sim -4K_1^2 \exp \xi_1 \cos \ze_1$, where 
$\xi_1= \Om_1 t + \Re (\eta_{10})$ and
$\ze_1= K_1 x + \Im (\eta_{10})$. 
Similarly, as $t \rightarrow  \infty$, 
$u \sim -4K_1^2/a_{12} \exp (-\xi_1) \cos \ze_1$. 
This behavior is no more than the linear approximation 
of the Boussinesq equation around $u=0$. 
Here $a_{12}$ plays a role of shift of time
and the offset of the amplitude. 
Figure 4 shows the 3D plot of the one-homoclinic solution. 

\begin{figure}[t]
\begin{center}
\mbox{
\epsfbox{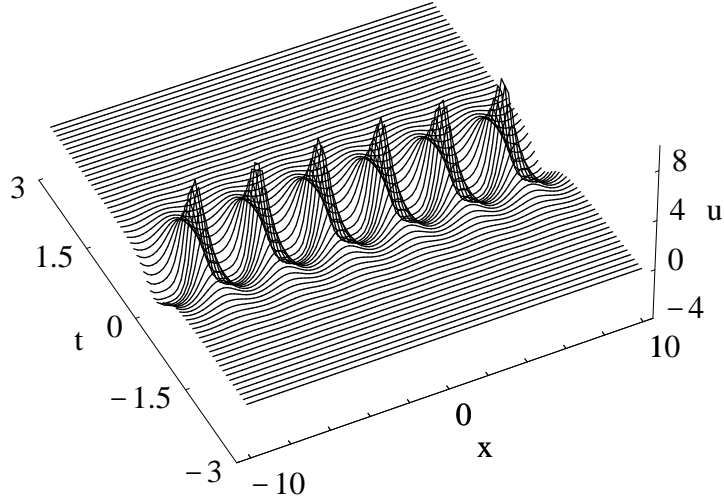}}
\end{center}
\caption{ The 3D plot of the one-homoclinic solution
of the Boussinesq equation with $k_{1} = 2i$ .}
\label{fig4}
\end{figure}

One remarkable feature of the Boussinesq equation 
with the spatial period $l_x$ comparing with the 
nonlinear Schr\"odinger equation \cite{Herbst} 
is that there is a countably infinite number of 
homoclinic orbits due to the short wavelength instability 
$k_j=iK_j$ with $K_j= 2\pi j/l_x>1 $. 

The necessary and sufficient condition for 
the regularity of the $(M,L)=(1,0)$ solution is $a_{12} > 1$. 
Using 
\beq
a_{12} = 
\frac{|1+K_1^2 \e^{2i\vph}|-1-4K_1^2\cos 2\vph_1+3K_1^2} 
{|1+K_1^2 \e^{2i\vph}|-1-4K_1^2\cos 2\vph_1-3K_1^2},
\label{a25} \eeq
and $K_1>1$, the condition can be summarized as 
\beq
\cos 2\vph_1 < -\fr{3}{4}-\fr{1}{4K_1^2}.
\label{a26} \eeq
For a complex wavenumber $k_1$, this solution 
may no longer be called homoclinic. 
Figure 5 shows the 3D plot of the solution 
with a complex wavenumber $k_1$ satisfying the condition 
(\ref{a26}). 

\begin{figure}[t]
\begin{center}
\mbox{
\epsfbox{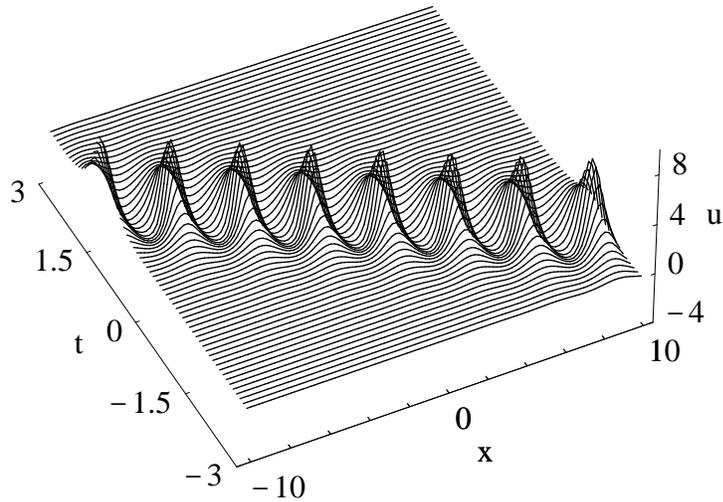}}
\end{center}
\caption{ The 3D plot of the one-wavepacket solution
of the Boussinesq equation with a complex wavenumber 
$k_{1} = 2\exp (i \vph_1) , \vph_1 = \cos ^{-1} (1/4) $ .}
\label{fig5}
\end{figure}

The solution shows that a wavepacket travelling at the 
group velocity $-\Om_{1r}/k_{1r}$ of waves with 
the phase velocity  $-\Om_{1r}/k_{1i}$. 
Therefore, it may be called a {\it one-wavepacket} solution. 
Similarly, we may construct two- and $N$-wavepacket solutions, 
and interactions between solitons, homoclinic solutions 
and wavepackets. 

We have shown that the Boussinesq equation has richer 
geometrical structures in phase space than the KdV equation, 
although the homoclinic solutions may be far from modelling 
shallow water waves because of the assumption 
$u_t \approx -u_x $ in the derivation of the equations 
\cite{Miles}.
The complexation of wavenumbers in solitons shown in 
this paper has a wide application and is expected to give 
a tool for finding a new class of exact solutions.


\begin{thebibliography}{99}
\bibitem{Herbst}
B. M. Herbst and M. J. Ablowitz, 
Phys. Rev. Lett. 62 (1989) 2065.

\bibitem{Umeki}
M. Umeki, RIMS kokyuroku 974 (1996)

\bibitem{Bishop}
A. R. Bishop, M. G. Forest, D. W. McLaughlin and 
E. A. Oberman, 
Phys. Lett A 126 (1988) 335.

\bibitem{Hirota1}
R. Hirota,
Phys. Rev. Lett. 27 (1971) 1192.

\bibitem{Hirota2}
R. Hirota,
J. Math. Phys. 14 (1973) 810.

\bibitem{Nakamura}
A. Nakamura, 
J. Phys. Soc. Jpn. 47 (1979) 1701.

\bibitem{Miles}
J. W. Miles
Ann. Rev. Fluid Mech. 12 (1980) 11.

\end{thebibliography}
\end{document}